\documentclass{article} % For LaTeX2e
\usepackage{iclr2021_conference,times}

\usepackage{amsmath,amsfonts,bm}

\def\eqref#1{equation~\ref{#1}}
\def\1{\bm{1}}

\DeclareMathAlphabet{\mathsfit}{\encodingdefault}{\sfdefault}{m}{sl}
\SetMathAlphabet{\mathsfit}{bold}{\encodingdefault}{\sfdefault}{bx}{n}

\usepackage{mathtools}

\usepackage{amssymb}

\usepackage{amsmath}
\usepackage{graphicx}

\usepackage{hyperref}
\usepackage{url}

\title{A Machine-Learning Surrogate Model for \textit{ab initio} Electronic Correlations at Extreme Conditions}

\author{Tobias Dornheim, Zhandos A. Moldabekov, \& Attila Cangi\\
Center for Advanced Systems Understanding (CASUS)\\
D-02826 G\"orlitz, Germany \\
\texttt{\{t.dornheim,z.moldabekov,a.cangi\}@hzdr.de} \\
}

\iclrfinalcopy % Uncomment for camera-ready version, but NOT for submission.
\begin{document}

\maketitle

\begin{abstract}
The electronic structure in matter under extreme conditions is a challenging complex system prevalent in astrophysical objects and highly relevant for technological applications. We show how machine-learning surrogates in terms of neural networks have a profound impact on the efficient modeling of matter under extreme conditions. We demonstrate the utility of a surrogate model that is trained on \emph{ab initio} quantum Monte Carlo data for various applications in the emerging field of warm dense matter research.
\end{abstract}

\section{Introduction}
Modeling complex systems is a complicated task. It requires taking into account hierarchies on different scales and their inherent nontrivial interactions. In Systems Biology, morphogenesis is an apt example. Phenomena on the nanoscale like chemical reactions of proteins determine the shape of an entire biological cell. Likewise, modeling air quality is a suitable example from the field of Earth System Research. Microscale meteorological phenomena like small atmospheric turbulences give rise to thunderstorms at the meteorological mesoscale. These, in turn, lead to atmospheric circulation phenomena on the global scale like El Nino. 

In this work, we focus on warm dense matter (WDM) which is a prime example of a hierarchical, complex system.
Induced by extreme electromagnetic fields, temperatures and pressures, WDM is a highly challenging phase of matter.
While WDM phenomena in state-of-the-art experiments take place on the atomic scale, their manifestations in astronomical objects like giant gas planet interiors~(see \cite{manuel}) and crusts of neutron stars have astronomical dimensions. For instance, electrons react on the level of picoseconds (i.e., $10^{-12}$s), whereas the evolution of stars and planets takes billions of years. These examples underline the fact that modeling WDM rigorously requires a bottom-up design that integrates computationally challenging \emph{ab initio} calculations on the atomic scale. Modeling such a complex system can be achieved by fusing digital twins of experiments with nano-physics informed models. 

From a physical perspective, WDM is defined by two characteristic parameters: (1) the density parameter $r_s=\overline{r}/a_\textnormal{B}$ with $\overline{r}$ denoting the mean electronic distance and $a_\textnormal{B}$ the first Bohr radius; (2) the degeneracy temperature $\theta = k_\textnormal{B}T/E_\textnormal{F}$ where $k_\textnormal{B}$ is the Boltzmann constant, $T$ the temperature, and $E_\textnormal{F}$ the Fermi energy (see \cite{quantum_theory} for details). 
The WDM regime is demarked by the region of parameter space where both characteristic parameters are simultaneously of the order of unity. This means that an accurate description of WDM on the atomic scale must take into account the intriguingly intricate interplay of a) the Coulomb repulsion between the electrons, which causes correlations amongst them, b) thermal excitations, which rule out ground-state methods used in quantum chemistry, and c) quantum degeneracy effects such as Pauli blocking, which prevents the use of semi-classical methods like molecular dynamics.
Due to these challenges, the arguably most promising option is to resort to \textit{ab initio} quantum Monte Carlo (QMC) methods (see e.g. \cite{anderson2007quantum,cep} for a detailed introduction). QMC methods provide a solution to the quantum many-body problem without any empirical input and are, in principle, exact. As a common denominator in all QMC methods, the physical property of interest (the total energy, pressure, etc.) is expressed as a high-dimensional integral, which is evaluated stochastically using the Monte Carlo method pioneered by \cite{metropolis}.

Yet, the QMC simulation of particles obeying Fermi-Dirac-statistics (\emph{fermions}, such as electrons) is severely hampered by the notorious fermion sign problem. It leads to an exponential scaling of the compute time with increasing system size or decreasing temperature; see \cite{dornheim_sign_problem} for a recent, accessible discussion of this issue. In fact, \cite{troyer} showed that the sign problem is $NP$-hard in many relevant cases, and that a general solution thus appears unlikely. This is a fundamental problem for modeling WDM, first and foremost, in support of WDM experiments. While an on-the-fly QMC simulation is beyond the scope even on modern supercomputers, many applications require a rapid evaluation of electronic properties on a continuous parameter grid (often mass density and temperature).

A first seminal step towards solving this problem has recently been presented by \cite{dornheim_ML}. A highly accurate surrogate model for electronic correlations covering the relevant range of parameters in WDM has been constructed by combining the output of different \textit{ab initio} techniques. In this work, we review this effort concisely and illustrate three of its applications: (1) supporting the interpretation of an X-Ray scattering experiment on warm dense aluminum by \cite{Sperling_PRL_2015}, (2) predicting yet unexplored nonlinear effects in WDM, and (3) computing energy-loss properties of WDM.

\section{A Surrogate Model for Electronic Correlations}

\begin{figure}[h]
\begin{center}
%\framebox[4.0in]{$\;$}
\includegraphics[width=0.463\textwidth]{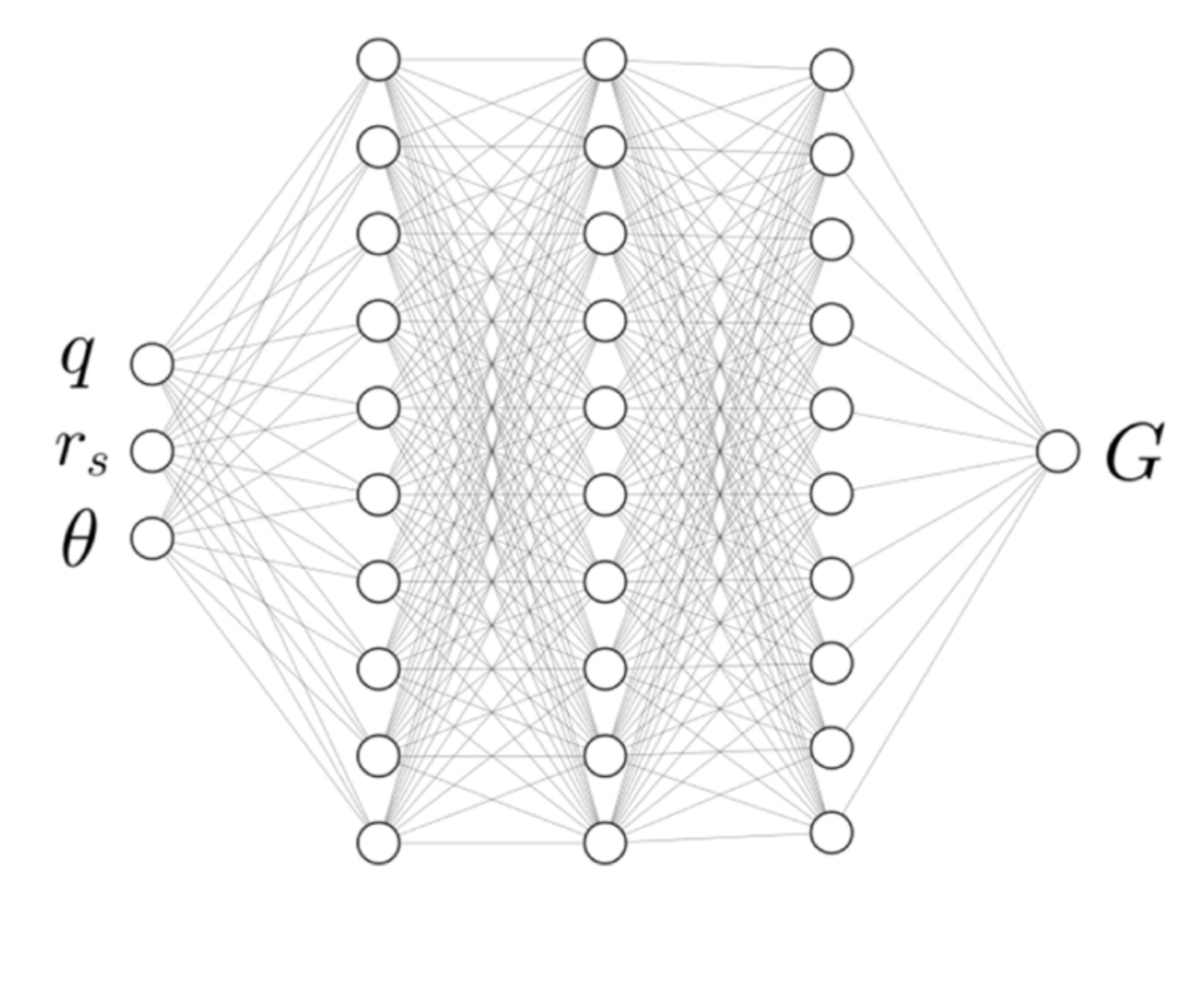}\hspace*{0.05\textwidth}\includegraphics[width=0.463\textwidth]{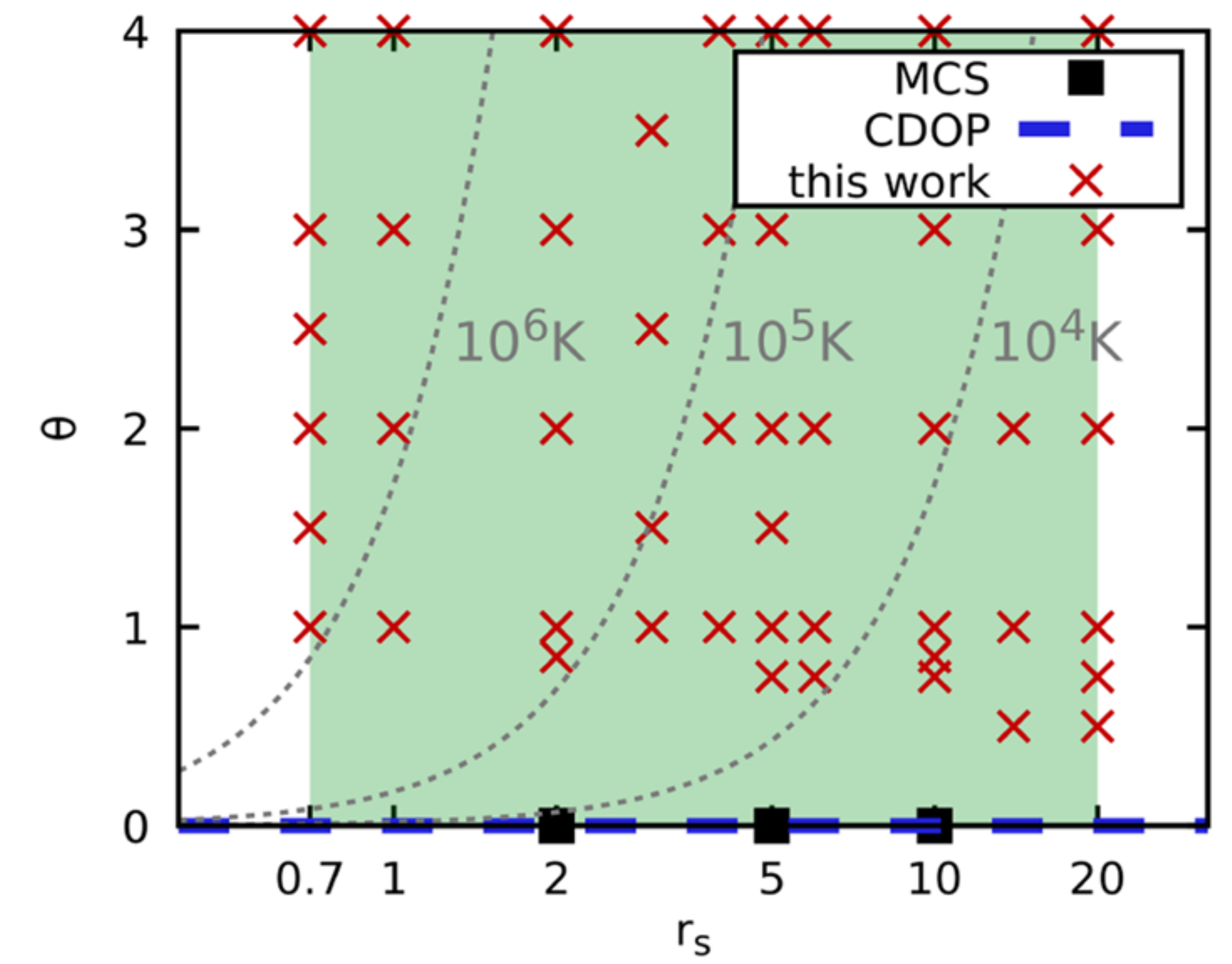}
\end{center}
\caption{\label{fig:basic}
Left: Schematic neural-network layout of the machine-learning surrogate model representing the static LFC. The actual network is comprised of $N_l=40$ fully connected hidden layers with $N_n=64$ neurons in each. Right: Density-temperature plane with available training data. MCS:~\cite{moroni2}; CDOP:~\cite{cdop}.  The neural net covers the shaded green region. Taken from \cite{dornheim_ML} with the permission of the authors.
}
\end{figure}

The central result of \cite{dornheim_ML} is a surrogate model for the electronic static local field correction (LFC) --- a central quantity in linear response theory that encodes electronic correlations in terms of wave-number resolved exchange--correlation effects; see e.g. the review by \cite{review}. This was achieved by training a fully connected deep neural network, which is valuable method for universal function approximation, e.g.~\cite{rolnick2018power}. The basic layout is shown in the left panel of Fig.~\ref{fig:basic}. The LFC $G(q;r_s,\theta)$ takes as input the wave-number $q$, the density parameter $r_s$, and the temperature parameter $\theta$. This tuple is then processed by the neural network. The output layer yields the corresponding $G$, which is a real number.

Naturally, this requires a sufficient amount of training data. This is illustrated in the right panel of Fig.~\ref{fig:basic}, which shows the parameters $r_s$ and $\theta$ where training data for $G$ are available (over the full $q$-range of interest). Available \textit{ab inito} path-integral Monte Carlo data at finite temperature ($\theta\gtrsim0.5$) are denoted by red crosses. These simulations quickly become infeasible at lower values of $\theta$ due to the exponential increase in compute time, i.e., the fermion sign problem discussed above.
In addition, \cite{moroni2} generated data for $G$ at $\theta=0$ using a complementary ground-state method, which is depicted by the black squares. Finally, these data were used by \cite{cdop} to obtain an analytical parametrization of $G(q;r_s,0)$, i.e., the dashed blue line at the bottom.

The task at hand is then two-fold: the neural-network needs to (1) reproduce the training data accurately and (2) interpolate smoothly between the provided data points. The details of the training procedure can be found in the original publication by \cite{dornheim_ML}, including the training loss, optimizer, and validation against independent data.
Here, we restrict ourselves to a brief discussion of the results, before we proceed to the applications of the surrogate model for the prediction of different material properties detailed in Sec.~\ref{sec:applications}.

\begin{figure}[h]
\begin{center}
%\framebox[4.0in]{$\;$}
\includegraphics[width=0.463\textwidth]{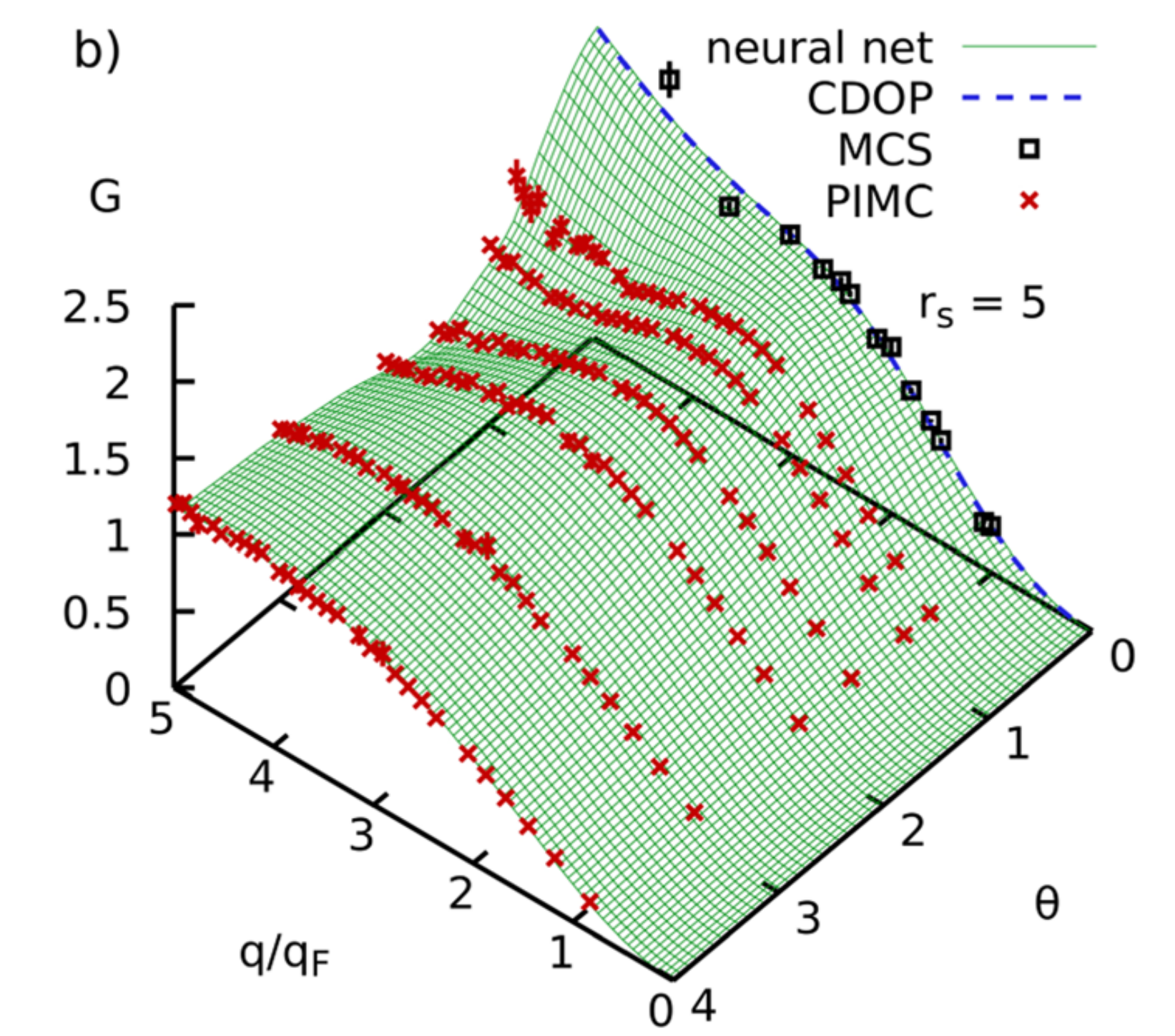}\hspace*{0.08\textwidth}\includegraphics[width=0.463\textwidth]{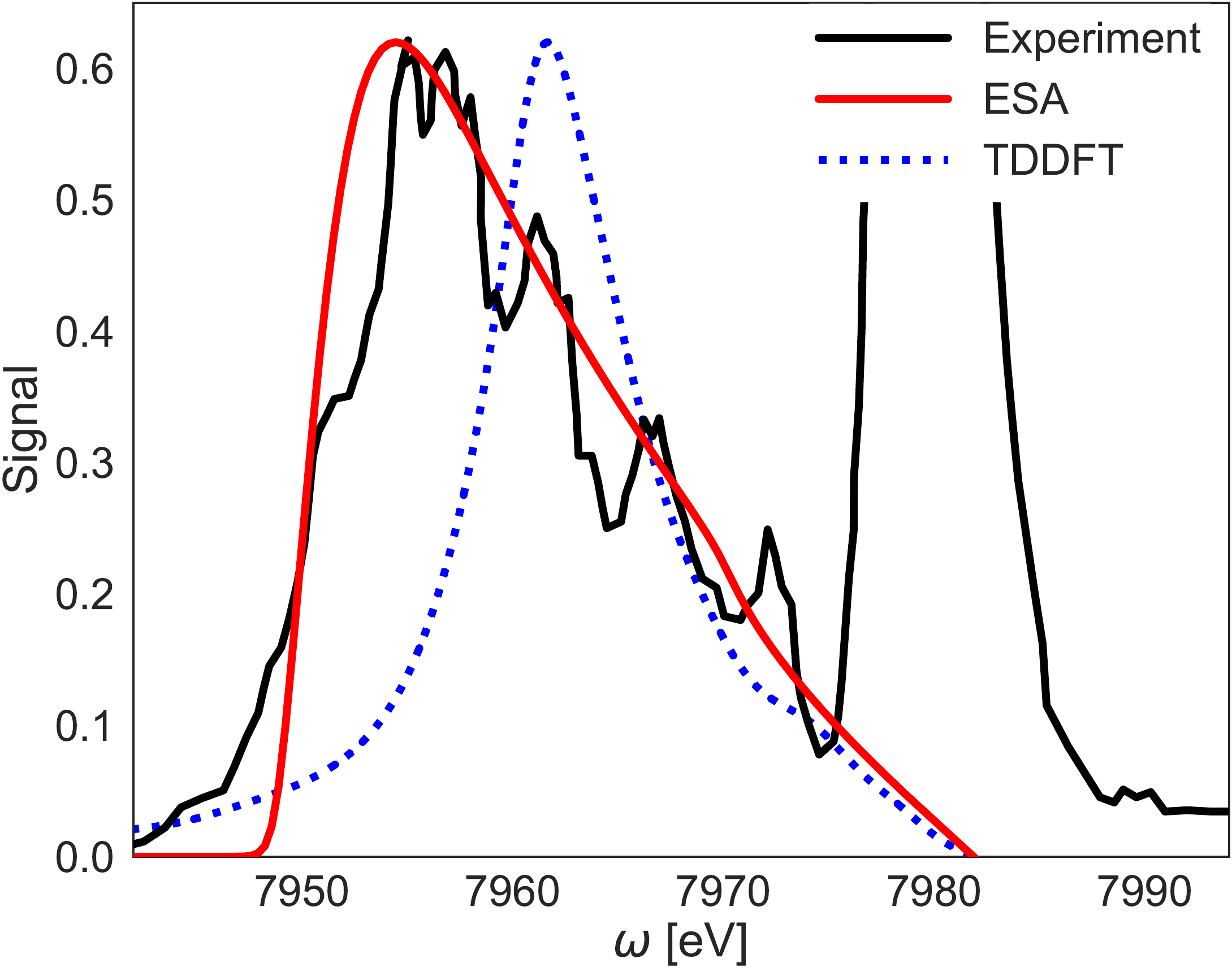}
\end{center}
\caption{\label{fig:net}
Left: Static LFC in the $\theta$-$q$-plain for $r_s=5$. The green represents the continuous surrogate model of the electronic correlations. Taken from~\cite{dornheim_ML} with the permission of the authors. Right: X-Ray Thomson scattering signal of aluminum: Experiment by~\cite{Sperling_PRL_2015}; red: prediction by \cite{Dornheim_PRL_2020_ESA} in terms of the surrogate model; dotted blue: time-dependent density functional theory.}
\end{figure}

In the left panel of Fig.~\ref{fig:net}, we show the LFC in the $q$-$\theta$-plane for a fixed value of the density parameter $r_s$. In particular, the red crosses show the path-integral Monte Carlo results at finite temperature and the black squares (dashed blue line) the QMC results (parametrization) in the ground state. Evidently, the trained neural net (green) is able to reproduce the training data to high accuracy and to interpolate smoothly between the provided data set. 
The surrogate model thus yields an \emph{ab initio} representation of electronic correlations on a continuous grid of parameters with negligible computational cost. Inference of the surrogate model takes seconds, whereas generating the training data took $\tau\sim\mathcal{O}\left(10^6\right)$ CPU hours on various HPC platforms.

\section{Applications\label{sec:applications}}
Let us now focus on the application of the surrogate model for different WDM applications. A particularly important use case is the interpretation of experiments. More specifically, the diagnostics of WDM experiments is challenging, because even basic system parameters like the temperature cannot be directly measured due to the extreme conditions. Instead, the system is probed by a free-electron X-Ray laser. The detected scattering signal is used to infer the plasma parameters of interest; see \cite{siegfried_review} for a review article on this method. An example for such a measured signal from an experiment on warm dense aluminum by \cite{Sperling_PRL_2015} is illustrated as the black curve in the right panel of Fig.~\ref{fig:net}. The task at hand is then to fit a theoretical prediction of the signal to the experimental data, which is an optimization problem in terms of the unknown plasma parameters, i.e., the temperature $T$.

A common and simple method is the \emph{random phase approximation} which neglects electronic exchange--correlation effects altogether. As shown, the resulting dashed yellow curve clearly does not reproduce the measured curve. A much more accurate and computationally more expensive method is time-dependent density functional theory (TDDFT) which takes about $\tau\sim\mathcal{O}\left(10^4\right)$ CPU hours on a computing cluster. Still, it only yields  qualitative agreement with the experimental signal.
Finally, we demonstrate the efficacy of our machine-learning surrogate model \cite{Dornheim_PRL_2020_ESA} which is illustrated as the solid red curve. Evaluating it takes only seconds. Evidently, the agreement with the experimental signal is striking. This opens up the possibility of an automated on-the-fly interpretation of such experiments at an accuracy unattainable with any other state-of-the-art method. A corresponding open-source code is currently being developed.

\begin{figure}[h]
\begin{center}
%\framebox[4.0in]{$\;$}
\includegraphics[width=0.463\textwidth]{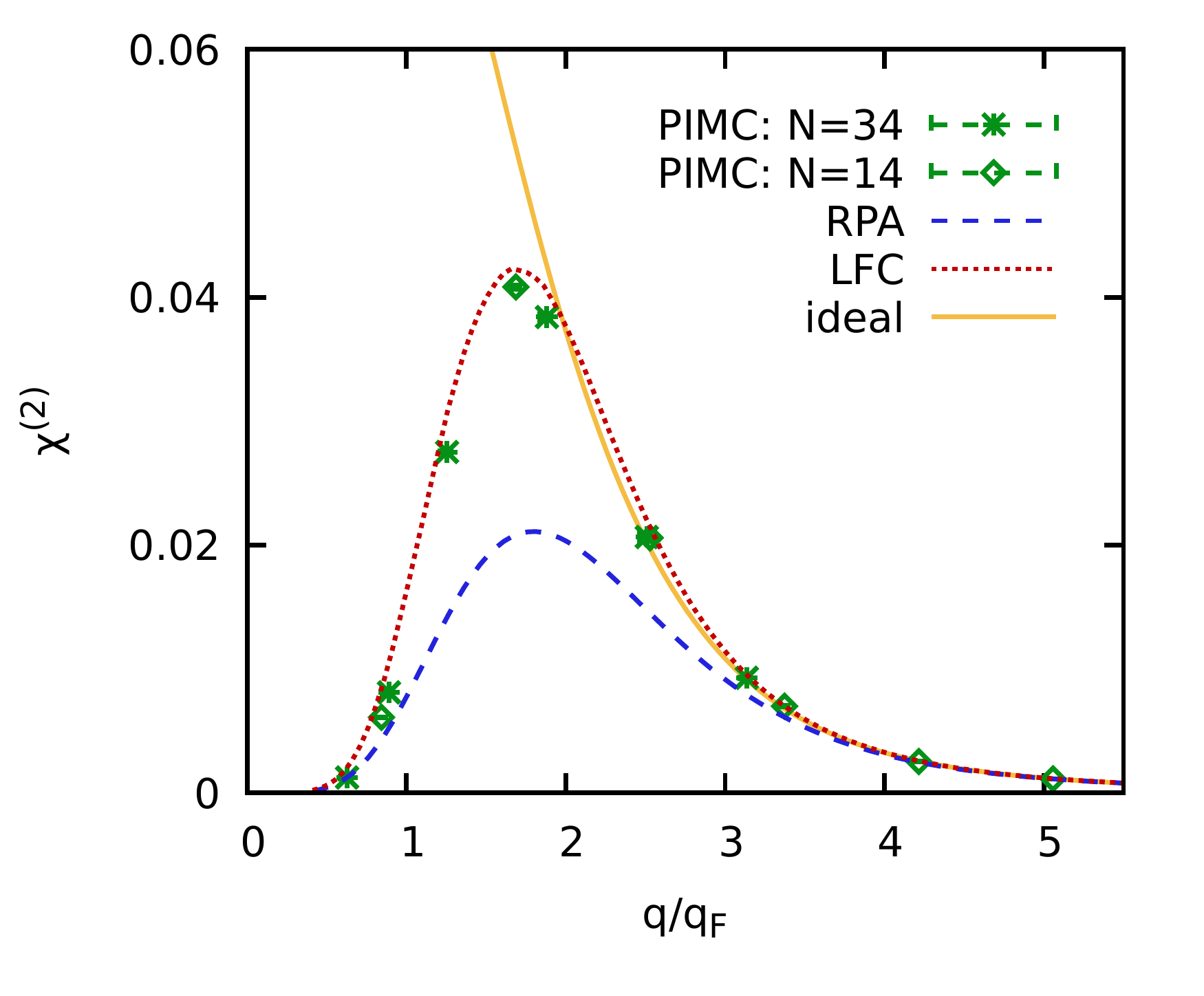}\hspace*{0.08\textwidth}\includegraphics[width=0.463\textwidth]{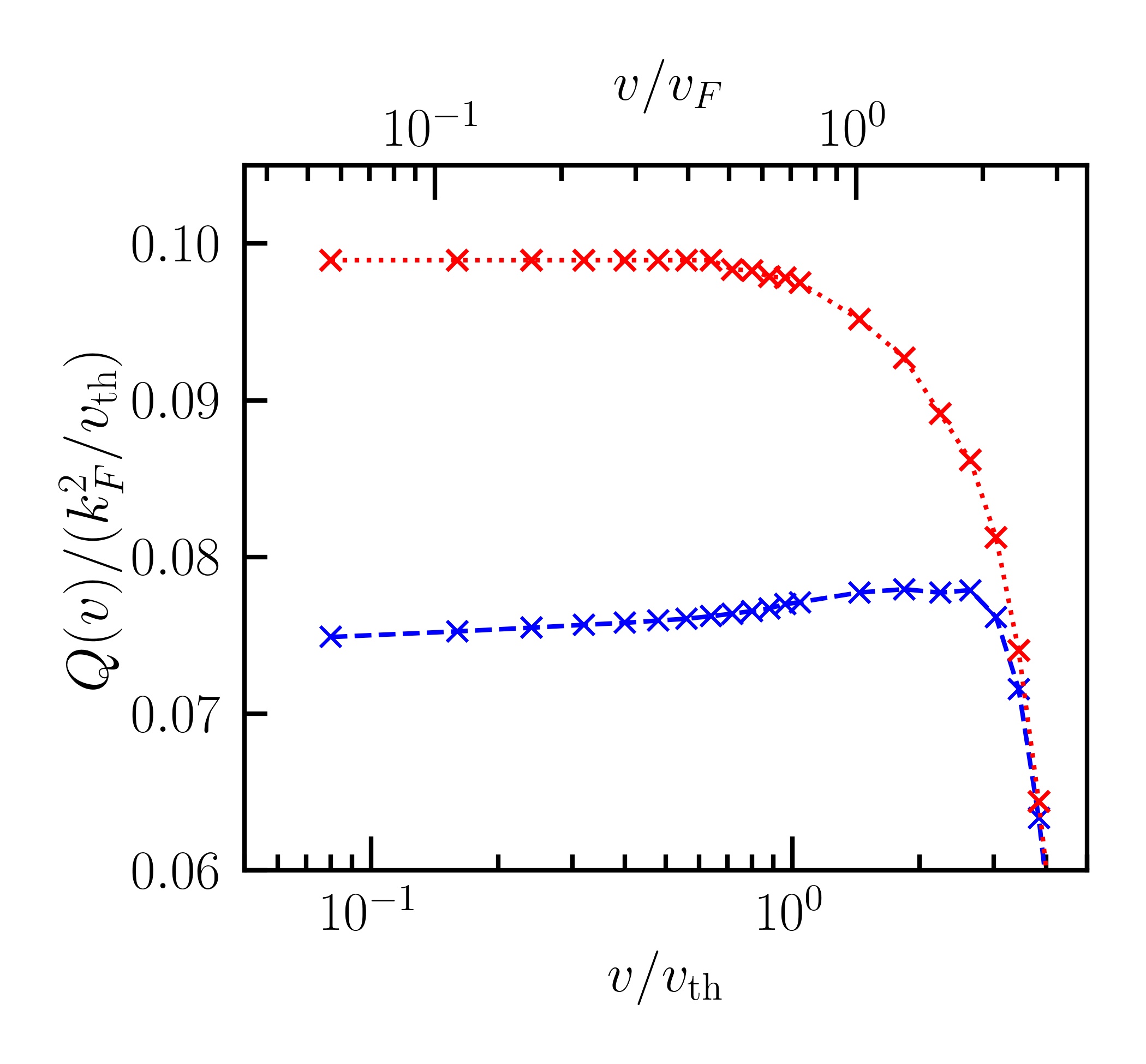}
\end{center}
\caption{\label{fig:app1}
Left: Quadratic density response function of the second harmonic; green: exact path-integral Monte Carlo (PIMC) results for $N=14$ and $N=34$ electrons; blue: RPA, red: surrogate-model, taken from Ref.~\cite{dornheim2021density}. Right: Electronic friction function, taken from~\cite{Moldabekov_PRE_2020}.
}
\end{figure}

A second example are nonlinear effects in the electronic density response of WDM, see \cite{Dornheim_PRL_2020,dornheim2021density} shown in the left panel of Fig.~\ref{fig:app1}. 
Exact path-integral Monte Carlo results for the excitation of the second harmonic of electrons in the WDM regime that are subject to a static perturbation of wave number $q$ are denoted as green stars and diamonds. Next, the prediction based on the random phase approximation discussed above is shown as the dashed blue line. However, it underestimates the magnitude by a factor of $1/2$. In contrast, our machine-learning surrogate model of $G(q;r_s,\theta)$ (dotted red curve) is in nearly perfect agreement with the exact data. We again stress that the calculation of the entire red curve took a few seconds, whereas computing a single data point using path-integral Monte Carlo takes $\tau\sim\mathcal{O}\left(10^4\right)$ CPU hours.

The final application is on the electronic friction function $Q(v)$, which is defined as the ratio between the stopping power of a target medium and the projectile velocity ~(\cite{Moldabekov_PRE_2020}). The friction function causes dissipation effects in the ion dynamics due to the energy exchange between electrons and ions --- an effect that is sensitive to the collective oscillation spectrum of the ions as was shown by \cite{Hanno_POP}. While important as a physical effect, taking it into account is computationally highly expensive due to the aforementioned complexities of WDM modeling. Therefore, the classical Rayleigh model (\cite{PhysRevE.77.061136}) is often used to approximate the friction in terms of free electrons (\cite{PhysRevLett.104.245001, Kang_2018}). This is a crude approximation that misses essential quantum and exchange-correlation effects. However, we can overcome the computational bottleneck and include quantum effects in terms of our machine-learning surrogate model. We, hence, precompute the friction function in the parameter range of interest efficiently in $\tau\sim\mathcal{O}\left(10 \right)$ CPU hours and then utilize it for running the Langevin dynamics of ions (\cite{Moldabekov_PRE_2020}). The results for the friction function at relevant WDM parameters ($\theta=0.5$ and $r_s=2$) are shown in the right panel of Fig.~\ref{fig:app1}. In comparison to the crude random phase approximation (blue), using the surrogate model (red) leads to a $30\%$ increase in the friction function at velocities order of thermal velocity. This clearly illustrates the importance of the exchange-correlation effects for the friction function in WDM.

\section{Summary and Outlook}
In this work, we summarized recent results by \cite{dornheim_ML} on a machine-learning surrogate model of electronic exchange--correlation effects at extreme densities and temperatures. The surrogate model yields highly accurate results at negligible computational cost. We presented three different applications of the surrogate model for the description of WDM. More specifically, we demonstrated the utility of the surrogate model for on-the-fly interpretation of experiments. Furthermore, we illustrated its ability to model nonlinear effects in WDM. Finally, we demonstrated how the surrogate model is utilized to compute energy-loss properties like the electronic friction and the stopping power in WDM.

We stress that we only presented a small selection of possible applications. Others include the construction of electronically screened potentials, the computation of thermal and electrical conductivities, the extension of quantum hydrodynamics, and the construction of advanced exchange--correlation functionals for density functional theory both in the ground state and at extreme conditions.

The construction of accurate surrogate models based on complex HPC data is a general approach. Potential prospective applications of this surrogate model include the extension to other electronic properties such as the momentum distribution and the thermal density matrix. 
As a final outcome, we envision constructing reliable digital twins of experiments. These would enable unprecedented insights into the interplay of different physical effects. Additionally, they would lead to the design of a new generation of experiments guided by rigorous theoretical descriptions with unassailable predictive capability.

\subsubsection*{Acknowledgments}
This work was partly funded by the Center for Advanced Systems Understanding (CASUS) which is financed by Germany's Federal Ministry of Education and Research (BMBF) and by the Saxon Ministry for Science, Culture and Tourism (SMWK) with tax funds on the basis of the budget approved by the Saxon State Parliament.
The PIMC calculations were carried out at the Norddeutscher Verbund f\"ur Hoch- und H\"ochstleistungsrechnen (HLRN) under grant shp00026, and on a Bull Cluster at the Center for Information Services and High Performace Computing (ZIH) at Technische Universit\"at Dresden.

\bibliography{bibliography}

\begin{thebibliography}{22}
\providecommand{\natexlab}[1]{#1}
\providecommand{\url}[1]{\texttt{#1}}
\expandafter\ifx\csname urlstyle\endcsname\relax
  \providecommand{\doi}[1]{doi: #1}\else
  \providecommand{\doi}{doi: \begingroup \urlstyle{rm}\Url}\fi

\bibitem[Anderson(2007)]{anderson2007quantum}
J.B. Anderson.
\newblock \emph{Quantum Monte Carlo: Origins, Development, Applications}.
\newblock Oxford University Press, USA, 2007.
\newblock ISBN 9780195310108.
\newblock URL \url{https://books.google.de/books?id=\_QUSDAAAQBAJ}.

\bibitem[Ceperley(1995)]{cep}
D.~M. Ceperley.
\newblock Path integrals in the theory of condensed helium.
\newblock \emph{Rev. Mod. Phys}, 67:\penalty0 279, 1995.
\newblock URL
  \url{https://journals.aps.org/rmp/abstract/10.1103/RevModPhys.67.279}.

\bibitem[Corradini et~al.(1998)Corradini, Sole, Onida, and Palummo]{cdop}
M.~Corradini, R.~Del Sole, G.~Onida, and M.~Palummo.
\newblock Analytical expressions for the local-field factor $g(q)$ and the
  exchange-correlation kernel ${K}_{\mathrm{xc}}(r)$ of the homogeneous
  electron gas.
\newblock \emph{Phys. Rev. B}, 57:\penalty0 14569, 1998.
\newblock URL \url{http://link.aps.org/doi/10.1103/PhysRevB.57.14569}.

\bibitem[Dai et~al.(2010)Dai, Hou, and Yuan]{PhysRevLett.104.245001}
Jiayu Dai, Yong Hou, and Jianmin Yuan.
\newblock Unified first principles description from warm dense matter to ideal
  ionized gas plasma: Electron-ion collisions induced friction.
\newblock \emph{Phys. Rev. Lett.}, 104:\penalty0 245001, Jun 2010.
\newblock \doi{10.1103/PhysRevLett.104.245001}.
\newblock URL \url{https://link.aps.org/doi/10.1103/PhysRevLett.104.245001}.

\bibitem[Dornheim(2019)]{dornheim_sign_problem}
T.~Dornheim.
\newblock Fermion sign problem in path integral {M}onte {C}arlo simulations:
  Quantum dots, ultracold atoms, and warm dense matter.
\newblock \emph{Phys. Rev. E}, 100:\penalty0 023307, 2019.
\newblock URL
  \url{https://journals.aps.org/pre/abstract/10.1103/PhysRevE.100.023307}.

\bibitem[Dornheim et~al.(2018)Dornheim, Groth, and Bonitz]{review}
T.~Dornheim, S.~Groth, and M.~Bonitz.
\newblock The uniform electron gas at warm dense matter conditions.
\newblock \emph{Phys. Reports}, 744:\penalty0 1--86, 2018.
\newblock URL
  \url{https://www.sciencedirect.com/science/article/abs/pii/S0370157318300516}.

\bibitem[Dornheim et~al.(2019)Dornheim, Vorberger, Groth, Hoffmann, Moldabekov,
  and Bonitz]{dornheim_ML}
T.~Dornheim, J.~Vorberger, S.~Groth, N.~Hoffmann, Zh.A. Moldabekov, and
  M.~Bonitz.
\newblock The static local field correction of the warm dense electron gas: An
  ab initio path integral {M}onte {C}arlo study and machine learning
  representation.
\newblock \emph{J. Chem. Phys}, 151:\penalty0 194104, 2019.
\newblock URL \url{https://aip.scitation.org/doi/full/10.1063/1.5123013}.

\bibitem[Dornheim et~al.(2020{\natexlab{a}})Dornheim, Cangi, Ramakrishna,
  B\"ohme, Tanaka, and Vorberger]{Dornheim_PRL_2020_ESA}
Tobias Dornheim, Attila Cangi, Kushal Ramakrishna, Maximilian B\"ohme,
  Shigenori Tanaka, and Jan Vorberger.
\newblock Effective static approximation: A fast and reliable tool for
  warm-dense matter theory.
\newblock \emph{Phys. Rev. Lett.}, 125:\penalty0 235001, Dec
  2020{\natexlab{a}}.
\newblock \doi{10.1103/PhysRevLett.125.235001}.
\newblock URL \url{https://link.aps.org/doi/10.1103/PhysRevLett.125.235001}.

\bibitem[Dornheim et~al.(2020{\natexlab{b}})Dornheim, Vorberger, and
  Bonitz]{Dornheim_PRL_2020}
Tobias Dornheim, Jan Vorberger, and Michael Bonitz.
\newblock Nonlinear electronic density response in warm dense matter.
\newblock \emph{Phys. Rev. Lett.}, 125:\penalty0 085001, Aug
  2020{\natexlab{b}}.
\newblock \doi{10.1103/PhysRevLett.125.085001}.
\newblock URL \url{https://link.aps.org/doi/10.1103/PhysRevLett.125.085001}.

\bibitem[Dornheim et~al.(2021)Dornheim, Böhme, Moldabekov, Vorberger, and
  Bonitz]{dornheim2021density}
Tobias Dornheim, Maximilian Böhme, Zhandos~A. Moldabekov, Jan Vorberger, and
  Michael Bonitz.
\newblock Density response of the warm dense electron gas beyond linear
  response theory: Excitation of harmonics, 2021.

\bibitem[Giuliani \& Vignale(2008)Giuliani and Vignale]{quantum_theory}
G.~Giuliani and G.~Vignale.
\newblock \emph{Quantum Theory of the Electron Liquid}.
\newblock Cambridge University Press, Cambridge, 2008.

\bibitem[Glenzer \& Redmer(2009)Glenzer and Redmer]{siegfried_review}
S.~H. Glenzer and R.~Redmer.
\newblock X-ray thomson scattering in high energy density plasmas.
\newblock \emph{Rev. Mod. Phys}, 81:\penalty0 1625, 2009.
\newblock URL
  \url{https://journals.aps.org/rmp/abstract/10.1103/RevModPhys.81.1625}.

\bibitem[K\"ahlert(2019)]{Hanno_POP}
Hanno K\"ahlert.
\newblock Dynamic structure factor of strongly coupled {Y}ukawa plasmas with
  dissipation.
\newblock \emph{Physics of Plasmas}, 26\penalty0 (6):\penalty0 063703, 2019.
\newblock \doi{10.1063/1.5099579}.
\newblock URL \url{https://doi.org/10.1063/1.5099579}.

\bibitem[Kang \& Dai(2018)Kang and Dai]{Kang_2018}
Dongdong Kang and Jiayu Dai.
\newblock Dynamic electron{\textendash}ion collisions and nuclear quantum
  effects in quantum simulation of warm dense matter.
\newblock \emph{Journal of Physics: Condensed Matter}, 30\penalty0
  (7):\penalty0 073002, jan 2018.
\newblock \doi{10.1088/1361-648x/aa9e29}.
\newblock URL \url{https://doi.org/10.1088/1361-648X/aa9e29}.

\bibitem[Metropolis et~al.(1953)Metropolis, Rosenbluth, Rosenbluth, Teller, and
  Teller]{metropolis}
Nicholas Metropolis, Arianna~W. Rosenbluth, Marshall~N. Rosenbluth, Augusta~H.
  Teller, and Edward Teller.
\newblock Equation of state calculations by fast computing machines.
\newblock \emph{The Journal of Chemical Physics}, 21\penalty0 (6):\penalty0
  1087--1092, 1953.
\newblock \doi{10.1063/1.1699114}.
\newblock URL \url{https://doi.org/10.1063/1.1699114}.

\bibitem[Moldabekov et~al.(2020)Moldabekov, Dornheim, Bonitz, and
  Ramazanov]{Moldabekov_PRE_2020}
Zh.~A. Moldabekov, T.~Dornheim, M.~Bonitz, and T.~S. Ramazanov.
\newblock Ion energy-loss characteristics and friction in a free-electron gas
  at warm dense matter and nonideal dense plasma conditions.
\newblock \emph{Phys. Rev. E}, 101:\penalty0 053203, May 2020.
\newblock \doi{10.1103/PhysRevE.101.053203}.
\newblock URL \url{https://link.aps.org/doi/10.1103/PhysRevE.101.053203}.

\bibitem[Moroni et~al.(1995)Moroni, Ceperley, and Senatore]{moroni2}
S.~Moroni, D.~M. Ceperley, and G.~Senatore.
\newblock Static response and local field factor of the electron gas.
\newblock \emph{Phys. Rev. Lett}, 75:\penalty0 689, 1995.
\newblock URL \url{http://link.aps.org/doi/10.1103/PhysRevLett.75.689}.

\bibitem[Plyukhin(2008)]{PhysRevE.77.061136}
A.~V. Plyukhin.
\newblock Generalized fokker-planck equation, brownian motion, and ergodicity.
\newblock \emph{Phys. Rev. E}, 77:\penalty0 061136, Jun 2008.
\newblock \doi{10.1103/PhysRevE.77.061136}.
\newblock URL \url{https://link.aps.org/doi/10.1103/PhysRevE.77.061136}.

\bibitem[Rolnick \& Tegmark(2018)Rolnick and Tegmark]{rolnick2018power}
David Rolnick and Max Tegmark.
\newblock The power of deeper networks for expressing natural functions, 2018.

\bibitem[Sch\"ottler \& Redmer(2018)Sch\"ottler and Redmer]{manuel}
M.~Sch\"ottler and R.~Redmer.
\newblock Ab initio calculation of the miscibility diagram for hydrogen-helium
  mixtures.
\newblock \emph{Phys. Rev. Lett}, 120:\penalty0 115703, 2018.
\newblock URL
  \url{https://journals.aps.org/prl/abstract/10.1103/PhysRevLett.120.115703}.

\bibitem[Sperling et~al.(2015)Sperling, Gamboa, Lee, Chung, Galtier,
  Omarbakiyeva, Reinholz, R\"opke, Zastrau, Hastings, Fletcher, and
  Glenzer]{Sperling_PRL_2015}
P.~Sperling, E.~J. Gamboa, H.~J. Lee, H.~K. Chung, E.~Galtier, Y.~Omarbakiyeva,
  H.~Reinholz, G.~R\"opke, U.~Zastrau, J.~Hastings, L.~B. Fletcher, and S.~H.
  Glenzer.
\newblock Free-electron x-ray laser measurements of collisional-damped plasmons
  in isochorically heated warm dense matter.
\newblock \emph{Phys. Rev. Lett.}, 115:\penalty0 115001, Sep 2015.
\newblock \doi{10.1103/PhysRevLett.115.115001}.
\newblock URL \url{https://link.aps.org/doi/10.1103/PhysRevLett.115.115001}.

\bibitem[Troyer \& Wiese(2005)Troyer and Wiese]{troyer}
M.~Troyer and U.~J. Wiese.
\newblock Computational complexity and fundamental limitations to fermionic
  quantum {M}onte {C}arlo simulations.
\newblock \emph{Phys. Rev. Lett}, 94:\penalty0 170201, 2005.
\newblock URL \url{http://link.aps.org/doi/10.1103/PhysRevLett.94.170201}.

\end{thebibliography}
\bibliographystyle{iclr2021_conference}

\end{document}